# Unveiling soliton booting dynamics in ultrafast fiber lasers

**Hong-Jie Chen[1], Meng Liu[1], Jian Yao[1], Song Hu[1], Jian-Bo He[1], Ai-Ping Luo[1], Zhi-Chao Luo[1,*], and Wen-Cheng Xu[1,*]**

*[1]Guangdong Provincial Key Laboratory of Nanophotonic Functional Materials and Devices & Guangzhou Key Laboratory for Special Fiber Photonic Devices and Applications, South China Normal University, Guangzhou, Guangdong 510006, China*

**Email: zcluo@scnu.edu.cn; xuwch@scnu.edu.cn*

**Ultrafast fiber lasers play important roles in many aspects of our life for their various applications in fields ranging from fundamental sciences to industrial purposes. Passive mode-locking technique is a key step to realize the ultrafast soliton fiber lasers. However, the booting dynamics of the soliton fiber lasers have not yet been well understood. Herein, we reveal the soliton buildup dynamics of ultrafast fiber lasers operating both in anomalous and net-normal dispersion regimes. Based on the advanced experimental methodologies of spatio-temporal reconstruction and dispersive Fourier transform (DFT), the soliton booting dynamics are analyzed in the time and spectral domains. It was found that the booting dynamics of conventional and dissipative solitons operating in the anomalous and net-normal dispersion regimes, respectively, are different from each other due to the different pulse shaping mechanisms. In particular, the spectral interference pattern with strong relaxation oscillation behavior was observed near the mode-locking transition for conventional soliton, while no relaxation oscillation of spectral pattern was obtained for dissipative soliton. We firstly revealed that the spectral pattern distributions are induced by the transient structured soliton formation during the pulse shaping from the noise background. The experimental results were verified by the theoretical simulations. The obtained results would provide a general guideline for understanding the soliton booting dynamics in ultrafast fiber lasers, and will prove to be fruitful to the various communities interested in solitons and fiber lasers.**

The ability of generating ultrashort pulses enables fast development of ultrafast science and technology, which thus, in turn, motivates laser scientists to search for high-performance ultrafast lasers with desirable features for practical applications[1-3]. The ultrashort pulses can be generated from the lasers by the principle of passive mode-locking technologies[4-7]. Generally, the passively mode-locked ultrafast solid-state lasers are considered to possess better performance than fiber lasers[8,9]. However, with the rapid developments of both the laser and optical fiber technologies, now the performance of modern ultrafast fiber lasers becomes to be comparable with the solid-state lasers[11-13]. Therefore, the ultrafast fiber lasers are regarded as the candidates of the next-generation ultrashort pulse sources because of their obvious advantages such as robust operation, flexible light path, and excellent heat dissipation.

After achieving the passive mode-locking operation in fiber lasers, the ultrashort pulses can be treated as the optical solitons[14]. Due to the high peak power of soliton pulse, the nonlinear effect experienced by the soliton in the fiber could result in the exhibition of abundant nonlinear dynamics by combining with the cavity parameter design. Therefore, in recent decades extensive efforts have been directed toward the investigations of soliton evolution and dynamics in mode-locked fiber lasers, such as the multi-soliton patterns[15-17], vector soliton[18-20] and dissipative soliton resonance[21-25]. As a fundamental but important nonlinear phenomenon of ultrafast fiber lasers, the buildup process of the passive mode-locking can be used to describe how to form a soliton in the laser systems. The buildup dynamics of the mode locked soliton can be defined as the stochastic processes initiated from noise fluctuations. Several investigations have been addressed to the detections of the soliton buildup process for ultrafast lasers. However, due to the lack of advanced measurement technologies, the detections of passive mode locking dynamics were generally restricted to the time domain with a low temporal resolution[26-28], which might result in the omission of some important details. To date, the mode locking buildup time and pulse evolution in a large timescale have been well understood. It was found that the buildup time is related to the intracavity pulse power, while the pulse will show the relaxation oscillation in the time domain, similar to the Q-switched mode-locking before the stable mode-locked pulse formation[28].

With the great advances in the detection technologies of ultrashort pulses, the opportunities to experimentally analyze the soliton dynamics, or more exactly, soliton transient dynamics was reopened for the ultrafast laser community[29,30]. As mentioned above, the soliton buildup process is a non-repetitive, stochastic process in ultrafast lasers, which is actually a transient nonlinear phenomenon. Therefore, resolving the soliton non-repetitive booting dynamics in both the time and spectral domains needs the high-speed real-time oscilloscope and real-time single-shot spectral measurement tools, respectively. To date, the state-of-art high-speed oscilloscope can resolve the picosecond pulse within a long record timescale of microseconds. However, due to the low scan rate of the commercially available optical spectrum analyzer (OSA), the measurement of soliton spectrum by the OSA is actually an average effect rather than a real-time one, causing that many experimental demonstrations of soliton transient spectral dynamics are hidden behind the theoretical predictions. Recently, the dispersive Fourier transform (DFT) was proposed to enable the mapping of single-shot spectra to the temporal waveforms so as to be captured by a high-speed real-time oscilloscope[31]. Therefore, the obstacle of real-time spectral measurement can be cleared by DFT method. By virtue of the real-time spectral measurements, several soliton transient dynamics have been observed in the ultrafast lasers, including rogue waves[32-34], soliton explosions[35,36], and phase evolution of the bound solitons[37-39]. In fact, the real-time spectral measurement by DFT method has also been applied to resolve the spectral dynamics in buildup process of ultrafast lasers[40,41]. Very recently, a detailed study of build-up of femtosecond Kerr-lens mode-locking (KLM) Ti:sapphire laser was reported. Several critical phenomena during the soliton buildup, such as birth of the broadband spectrum and transient interference among the random multiple picosecond pulses, have been directly observed[41].

As we know, KLM Ti:sapphire lasers are typically not self-starting ones, whose mode-locking operation needs to be triggered by the external perturbations[42]. Comparing to the KLM Ti:sapphire laser, the fiber lasers, however, could always achieve the self-starting operation as long as the cavity conditions are properly set[43,44]. The different onset conditions of the passive mode locking between KLM Ti:sapphire lasers and ultrafast fiber lasers are originated from the different physical mechanisms of pulse formation in the cavities. In addition, in fiber lasers the propagating lightwave is confined in a small mode area of single mode fiber, more nonlinear effects will be experienced by the mode-locked soliton. Therefore, a comprehensive investigation of the soliton booting dynamics in ultrafast fiber lasers is needed for the purpose of better understanding the formation dynamics of soliton.

In this work, we will address this issue. The soliton buildup processes were investigated experimentally and theoretically in the ultrafast fiber lasers both in the anomalous and net-normal dispersion regimes. Due to the pulse shaping from the noise background to stable mode locking operation, the evolved pulses oscillated periodically with structured temporal profiles, which results in the formations of the transient structural spectral patterns. Specifically, the mode-locking spectra of the conventional soliton in the anomalous dispersion regime show strong relaxation oscillation in the central part of mode-locked soliton, while no spectral relaxation oscillations were found for dissipative soliton in the normal dispersion regime and the transient shock waves[45] have been observed. It was found that the different spectral dynamics were caused by the different pulse formation mechanisms for the two types of solitons operating in both dispersion regimes. The obtained results would deepen our understanding for the soliton booting dynamics in ultrafast fiber lasers, which will be beneficial for both the nonlinear optics and ultrafast laser communities.

# Results

## Experimental setup

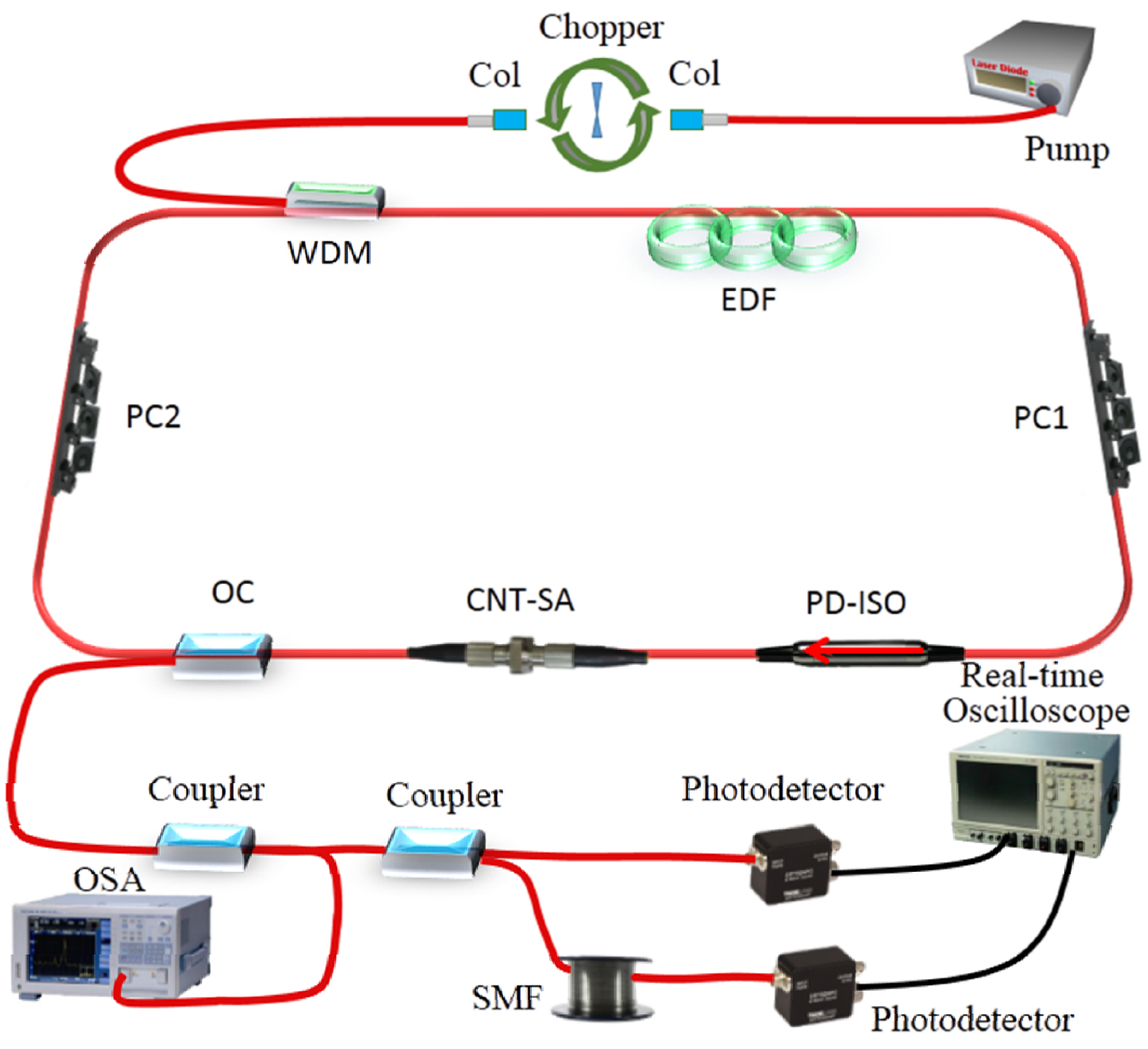

Figure 1. Ultrafast fiber laser used for investigating the soliton booting dynamics.

The ultrafast fiber laser used for investigating the soliton booting dynamics is shown in Fig. 1.

The erbium-doped fiber (EDF) was employed as the gain medium for the fiber laser operating at 1.55 μm waveband. A mechanical chopper is placed between the pump laser and the WDM to initiate (or stop) the self-starting mode locking operation. Two polarization controllers (PCs) were employed to adjust the polarization states. In order to effectively initiate the passive mode-locking operation, the carbon nanotube (CNT) was incorporated into the laser cavity acting as the saturable absorber (SA)[46,47]. For the purpose of better optimization of the mode locking status, we inserted a polarization-dependent isolator (PD-ISO) so as to tune the laser mode-locking parameters by the PCs, such as central wavelength and spectral bandwidth[48,49]. Therefore, the fiber laser was mode-locked by the hybrid mode-locking technique with the combination of nonlinear polarization rotation and CNT. To measure the soliton booting dynamics both in time and spectral domains with DFT method simultaneously, the output laser was divided by an additional 10:90 coupler. One port was directly connected to the high-speed real-time oscilloscope (Tektronix DSA-70804, 8 GHz) following by a photodetector (Newport 818-BB-35F, 12.5 GHz), while the other was linking to a dispersive element (~14 km long SMF) to map the mode-locked spectra into a temporal waveform directly shown on the oscilloscope by DFT[31,50]. As indicated above, two types of solitons operating in the anomalous and net-normal dispersion regimes were investigated. Therefore, for the conventional soliton in anomalous dispersion regime, the laser cavity was constructed by 4.1 m EDF with a dispersion parameter of 7.5 ps/nm/km and 11.5 m standard SMF. While for the dissipative soliton in normal dispersion regime, the cavity dispersion was compensated by a 10 m EDF with a dispersion parameter of -44.5 ps/nm/km and the length of the other SMF was 12 m.

**Conventional soliton booting dynamics in the ultrafast fiber laser with anomalous dispersion regime**

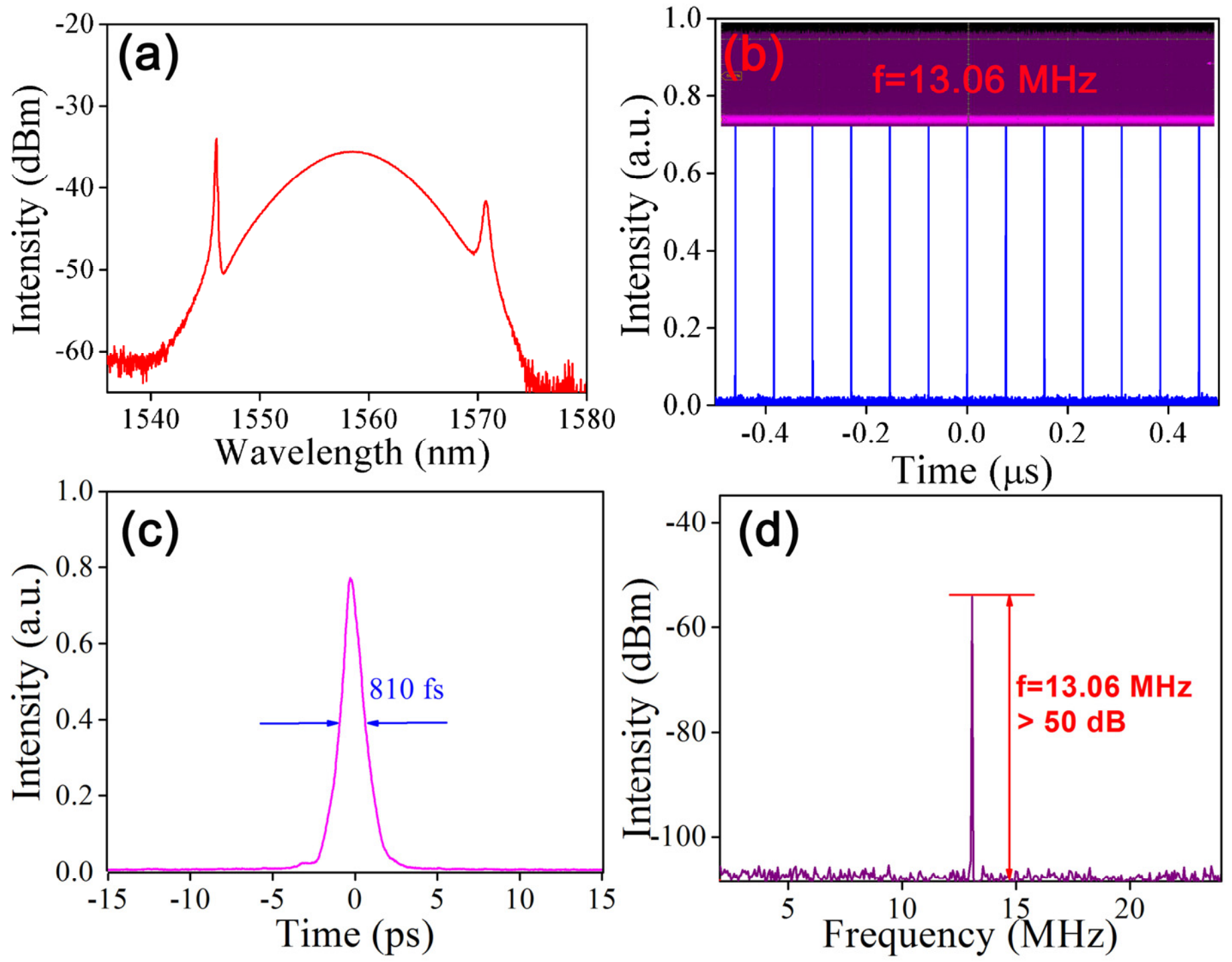

Figure 2. Typical performance of the conventional soliton. (a) Spectrum. (b) Pulse-train. Inset: pulse-train with larger range. (c) Autocorrelation trace. (d) RF spectrum.

Firstly, the conventional soliton booting dynamics in the ultrafast fiber laser with anomalous dispersion regime was discussed. As the CNT SA was incorporated in the laser cavity, the self-starting mode-locked operation of the fiber laser could be efficiently achieved at a pump power of 13 mW. We further increased the pump power to 13.6 mW and finely rotated the PCs to optimize the mode-locking operation. However, here we should note that the pump power needs to be restricted to a level which can ensure the fiber laser start in the single pulse regime. The typical performance of the ultrafast fiber laser was summarized in Fig. 2. The mode-locked spectrum centered at the wavelength of 1558 nm, with a 3-dB spectral bandwidth of 10.03 nm (Fig. 2(a)). The mode-locked pulse-train with a fundamental repetition rate of 13.06 MHz shows that the fiber laser operated in a single pulse regime. The pulse duration, which was characterized by a commercial autocorrelator (FR-103XL), is 810 fs, as can be seen in Fig. 2(c). In addition, the corresponding RF spectrum was measured in order to further check the stability of the ultrafast fiber laser. As shown in Fig. 2(d), the signal-to-noise ratio of RF spectrum was ~55 dB, indicating the high stability of the passive mode-locking operation.

To investigate the soliton booting process in the ultrafast fiber laser, then we opened the mechanical chopper to initiate/stop the mode locking operation periodically. Thus, the soliton buildup dynamics could be easily measured in this way. As mentioned above, the real-time spectral dynamics of the passive mode locking can be obtained by the DFT. Therefore, in our setup the soliton booting dynamics both in the time and spectral domains can be captured simultaneously by a single oscilloscope. Here, it should be noted that the temporal dynamics of the soliton booting from the noise background shows the obvious relaxation oscillations in the timescale of a few microseconds, which are similar to the previous reports[28] (see Supplementary Information). However, in this work we only concentrated on the dynamics very near the mode-locking transition in the fiber lasers. Since the pulse train and the spectral dynamics with DFT were all displayed on the oscilloscope as the time-continuous data stream, we segment the data stream into an interval of time length of 77.9 ns (cavity roundtrip time) to reconstruct the spatio-temporal and spatio-spectral dynamics of the passive mode locking across the consecutive roundtrips. In this way, the overview of the conventional soliton booting dynamics in time and spectral domains were plotted in Fig. 3. As can be seen from Fig. 3(a), the fiber laser operated from a narrow bandwidth quasi-continuous-wave (cw) to a mode-locking state with broadband spectrum. Notably, during the transition of the mode-locking process, the real-time spectral dynamics featured that the strong spectral relaxation oscillation was observed (also see the Supplementary Movie). Initially, the duration of the spectral relaxation oscillation could last for about 200 roundtrip time. Moreover, the period of the spectral relaxation oscillation possesses a decreasing trend when approaching the stable mode locking. Note that the duration of the spectral relaxation oscillation is related to the initial cavity parameter settings for stable mode-locking operation, which is case by case in our experiments. However, the spectral relaxation oscillation behavior could be always observed in the pre-mode-locking status of an ultrafast fiber laser operating in the anomalous dispersion regime. Correspondingly, the spatio-temporal dynamics depicted in Fig. 3(b) show that a continuous increase in the pulse intensity for the initial pre-mode-locking state, then experiences a decreasing trend and finally evolves to a stable mode-locking pulse-train. To more clearly describe this phenomenon, we provided the side view of Fig. 3(b) in Fig. 3(c). During the pulse evolution from noise to stable mode locking, only single pulse was detected on the oscilloscope traces, demonstrating that the soliton booting up dynamics are different from the KLM Ti:sapphire lasers[41]. It should be noted that the experimentally measured pulse evolution shown in Fig. 3(b) could not be fully resolved due to the limited bandwidth of the oscilloscope used. Then the measured soliton booting dynamics from the oscilloscope could be calculated to be the pulse energy evolution. To this end, we integrated the conventional soliton profiles from quasi-cw to mode-locking state for providing the pulse energy evolution in Fig. 3(d). In this case, the pulse energy firstly increased and then experienced an intensity drop before evolving into the stable mode-locking, possessing a similar trend to the spatio-temporal dynamics in Fig. 3(c).

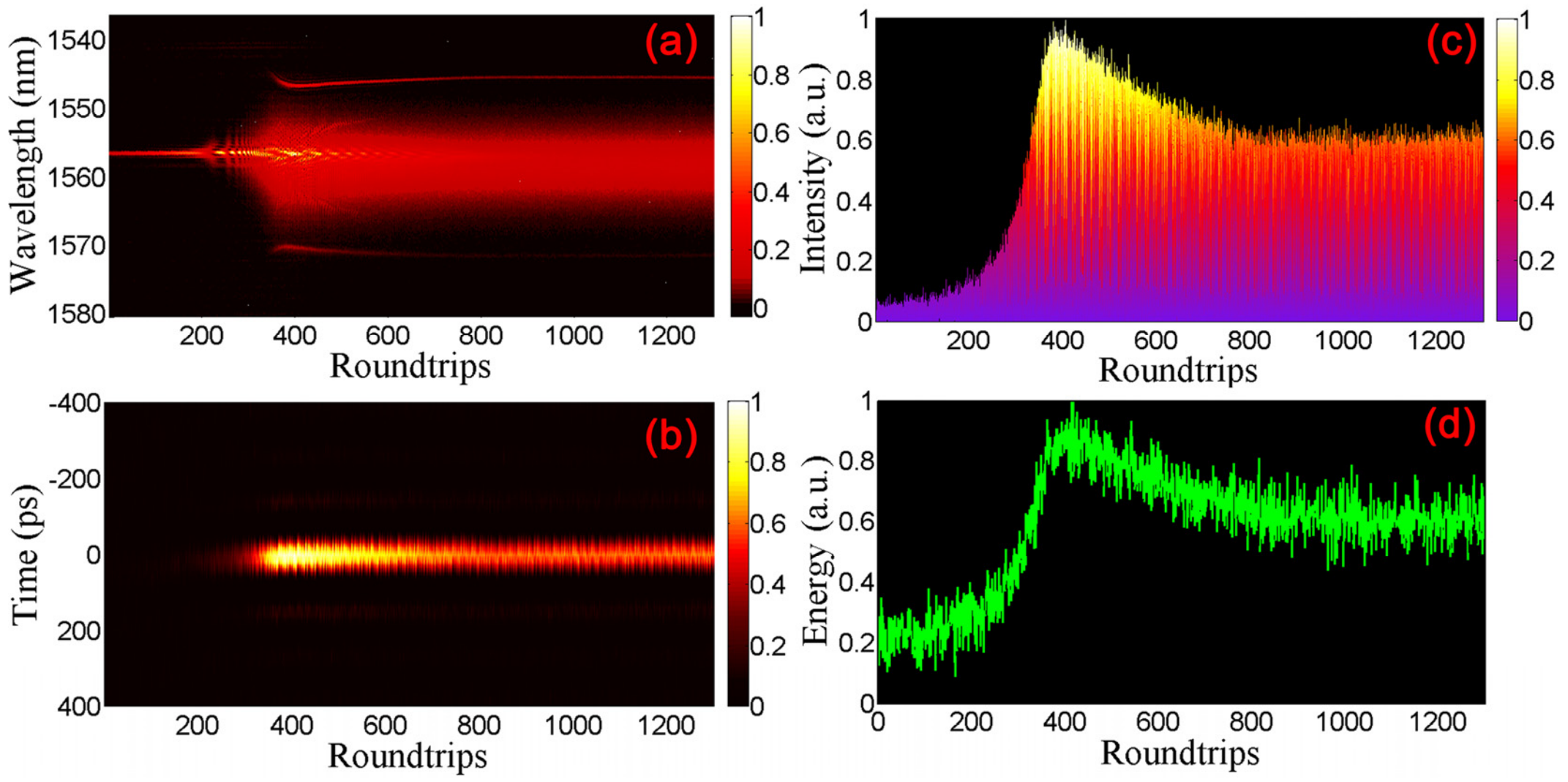

Figure 3. Conventional soliton booting dynamics in spectral and time domains for 1300 consecutive roundtrips. (a) Spatio-spectral dynamics. (b) Spatio-temporal dynamics. (c) Side view of spatio-temporal dynamics. (d) Pulse energy evolution.

According to the spectral dynamics in the pre-mode-locking state, we suspect that there will be some fine structures for the spectra as well as for the corresponding solitons in the very near the mode-locking transition state. To more clearly analyze the conventional soliton booting dynamics in ultrafast fiber lasers, we provided four transient spectral profiles of the booting soliton in Fig. 4 corresponding to different roundtrips. It can be seen that the spectral peaks or dips were alternatively generated in the central part of the mode-locked spectrum. In addition, the spectral interference patterns were also found. Based on the spectral dynamics shown in Fig. 4, we believed that the transient structural solitons[51] were formed during the soliton shaping to stable mode locking by saturable absorption effect. A specific structured soliton corresponds to a specific spectral profile during the soliton booting process[51]. Note that the structural soliton is not constant but evolves during the booting dynamics because of the strong pulse shaping in the laser cavity. Thus, the strong spectral relaxation oscillation could be seen in this stage until the stable soliton was finally formed. It should be also noted that in our temporal pulse evolution the structural profiles of the pulses could not be resolved experimentally. Nevertheless, this issue will be confirmed and discussed in the numerical simulation section.

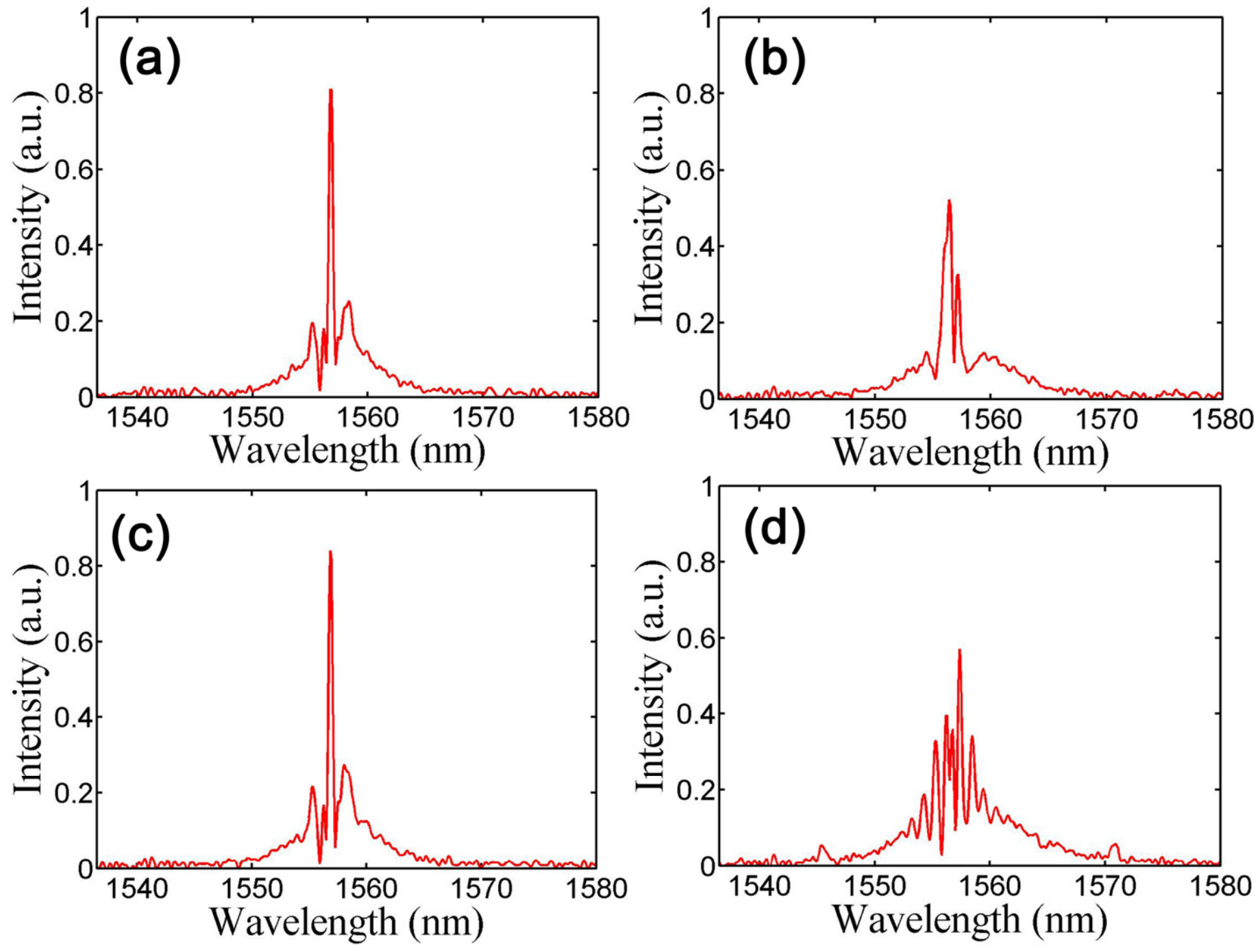

Figure 4. Four transient spectral profiles of the booting conventional soliton. (a) 324 roundtrip. (b) 326 roundtrip. (c) 328 roundtrip. (d) 356 roundtrip.

**Dissipative soliton booting dynamics in the net-normal dispersion fiber laser**

In the following, the dissipative soliton booting dynamics in the net-normal dispersion fiber laser will be investigated. Similar to the case of the conventional soliton, the self-started mode-locking could be easily achieved by increasing the pump power to 12.7 mW as the cavity parameters were properly adjusted. The laser performance of the dissipative soliton is displayed in Fig. 5. The typical rectangular spectral profile of the dissipative soliton was obtained with a center wavelength of 1562 nm[52,53]. The pulse-train operated in a fundamental repetition rate of 9.27 MHz and the pulse duration is 16.21 ps. Moreover, the RF spectrum shows a signal-to-noise ratio of ~60 dB, presenting that the dissipative soliton fiber laser operates in a relatively stable mode.

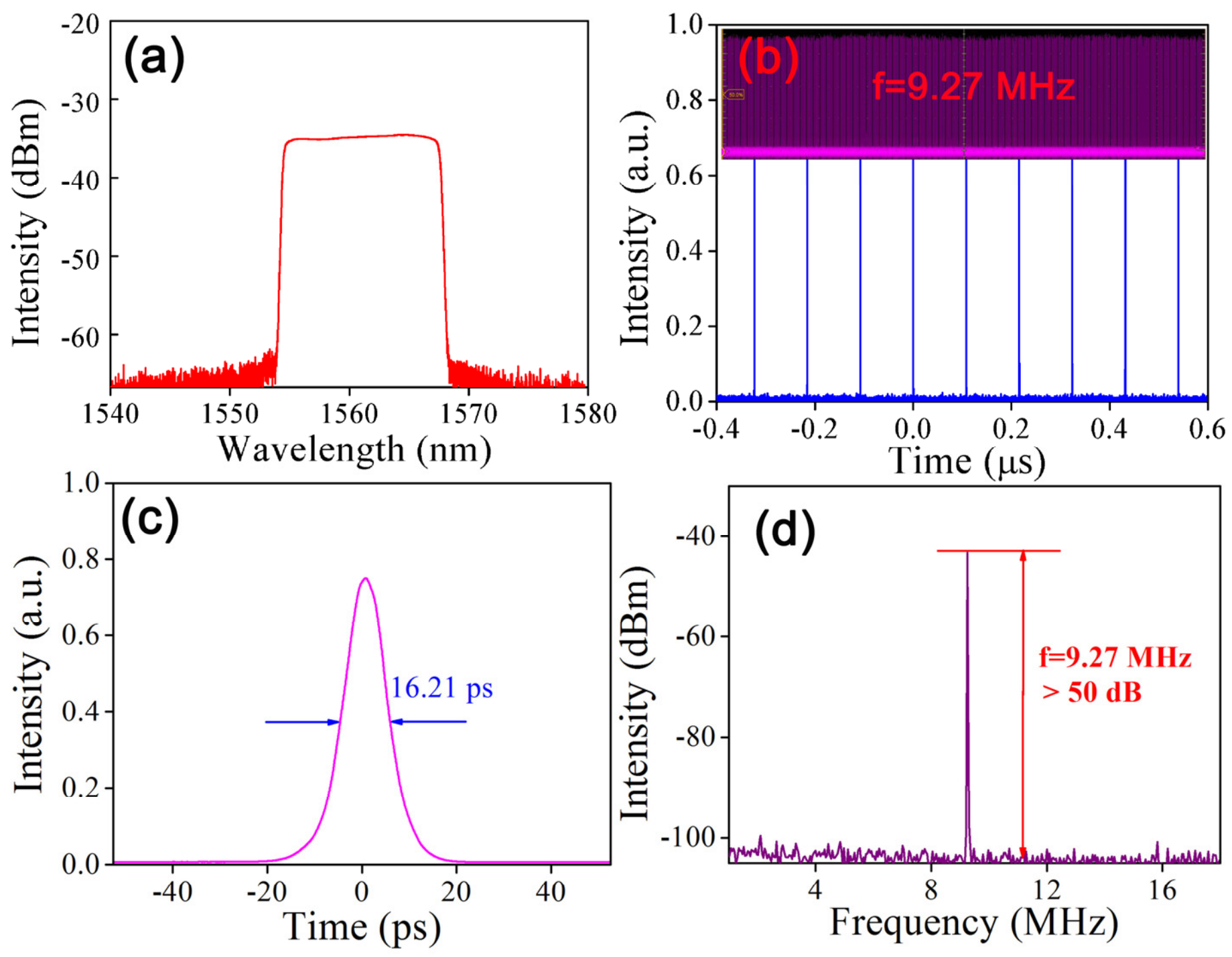

Figure 5. Typical performance of dissipative soliton. (a) Spectrum. (b) Pulse-train. Inset: pulse-train with larger range. (c) Autocorrelation trace. (d) RF spectrum.

Again, by periodically chopping the 980 nm pumping laser when the pump power was a little higher than the mode-locking threshold, the transition from cw to single-soliton mode-locked operation could be easily obtained. By virtue of DFT, the spatio-spectral dynamics of the dissipative soliton mode-locking process for 2700 consecutive roundtrips could be recorded by the real-time oscilloscope, as shown in Fig. 6(a). Similar to the case of conventional soliton booting dynamics, the spectral broadening process could be seen in the formation of dissipative soliton in the fiber laser. However, it is worth noting that the fine details of the spectral evolution in normal dispersion regime are different in comparison to those in the anomalous dispersion. The differences are manifested by the fact that the highest structured spectral spikes spread along both edges for dissipative soliton during the soliton buildup (also see the Supplementary Movie), while the strongest oscillating spectral components are always concentrated in the central part of conventional soliton during the booting process in the anomalous dispersion regime. Here, it should be also noted that, unlike the anomalous dispersion regime, no spectral relaxation oscillations were found. As for the spatio-temporal dynamics of the dissipative soliton build up process, it can be seen from the Fig. 6(b) and 6(c) that the pulse also evolved with an increasing trend in the peak intensity, then decrease to a certain value and finally become stable mode-locking state in fiber laser, which is similar to the booting dynamics of conventional soliton. However, here the decreasing amount of the pulse intensity before achieving the stable mode locking is much larger than that of the conventional soliton. In addition, Fig. 6 (d) also provided the pulse energy evolution in dissipative soliton booting regime, possessing similar evolution trend of the pulse-train envelope shown in Fig. 6 (c).

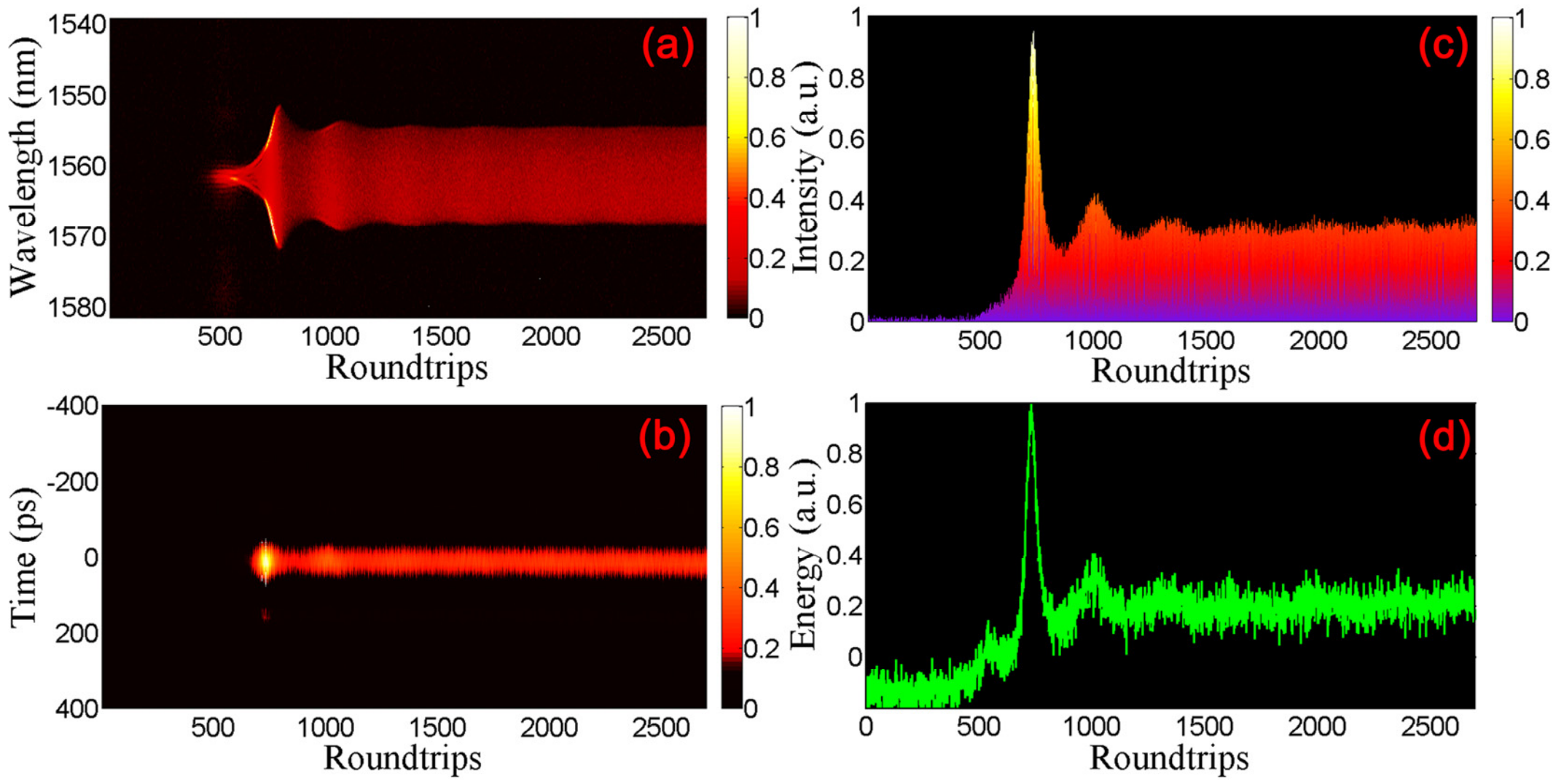

Figure 6. Dissipative soliton booting dynamics in spectral and time domains for 2700 consecutive roundtrips. (a) Spatio-spectral dynamics. (b) Spatio-temporal dynamics. (c) Side view of spatio-temporal dynamics (d) Pulse energy evolution.

The way to analyze the dissipative soliton booting dynamics is the same as that of the conventional soliton in anomalous dispersion regime. To gain insight into the transient dynamics of dissipative soliton in the fiber laser, we also provided four spectral profiles of the booting dissipative soliton corresponding to different roundtrips in Fig. 7. From Fig. 7, it can be seen that the spectral bandwidth becomes larger and larger when evolving into the stable mode-locking state. Meanwhile, the spectra also exhibit the structured profiles, where the highest interference peaks generally locate at the two edges of the evolved spectrum. Specifically, two sharp spectral peaks with oscillation structures around them could be seen on both edges of the pulse spectrum near the stable mode-locking operation, namely, the transient shock waves have been observed[45], as shown in Fig. 7(c). Following the generation of shock waves, after a certain number of roundtrips the bandwidth of the spectrum still broadens slowly until reaching the stable mode-locking state. In this case, a rectangular shape spectrum which is a typical characteristic of the dissipative soliton in fiber laser could be seen[52,53], indicating that the fiber laser operated in a stable mode-locking state, as presented in Fig. 7(d).

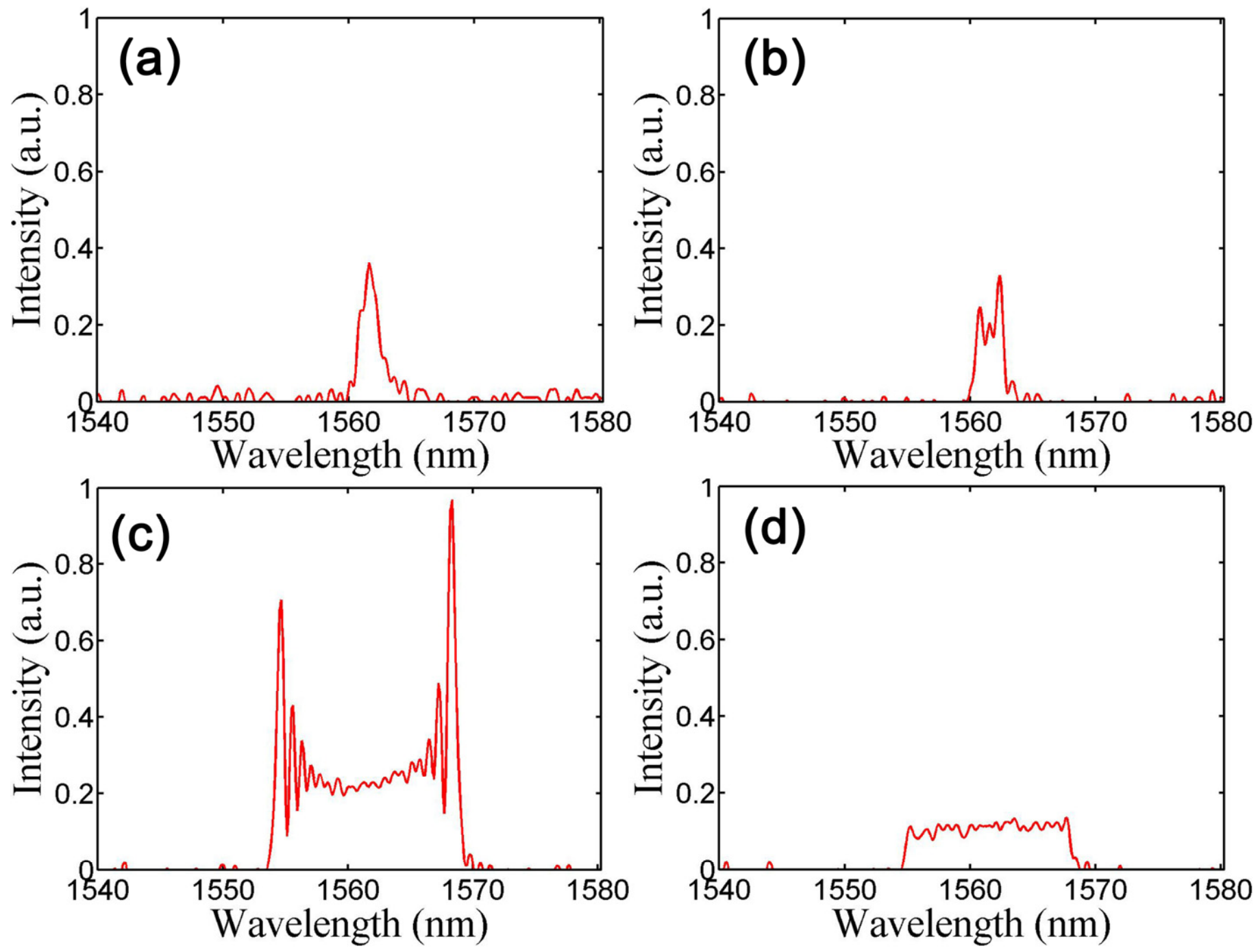

Figure 7. Four transient spectral profiles of the booting dissipative soliton. (a) 535 roundtrip. (b) 602 roundtrip. (c) 736 roundtrip. (d) 965 roundtrip.

**Numerical simulations of conventional soliton booting dynamics in ultrafast fiber laser**

As discussed above, the booting dynamics of the passive mode locking in fiber lasers operating in both the anomalous and normal dispersion regimes are experimentally observed from the spatio-spectral and spatio-temporal views. The following discussions are devoted to confirm these results by numerical simulations of the spectrum and pulse evolutions. The detailed simulation model and the parameters can be found in the supplementary file. And the simulation parameters are reasonably set according to the experimental conditions. In the following, we firstly presented the simulation results of the ultrafast fiber laser operating in the anomalous dispersion regime. The self-started mode-locking operation could be always seen in the simulation model of the fiber laser if the parameters were properly set. From the spectral evolution shown in Fig. 8(a), it can be seen that the fiber laser starts from quasi-cw with a narrow spectral bandwidth to the stable mode-locking spectrum with Kelly sidebands in the anomalous dispersion regime. Moreover, before achieving the stable mode-locking state, the spectral oscillation behavior could be also observed evidently. And the oscillation time period decreased when approaching the stable mode-locking state, where the trend is the same as that of experimental results. The corresponding pulse evolution from quasi-cw to mode locking in the time domain was plotted in Fig. 8(b) (top view) and (c) (side view). Here, the simulation pulse evolution shows that the pulse intensity increased firstly, then evolved into a stable one in the laser cavity. Figure 8(d) displays the integrated pulse energy evolution of simulation soliton from quasi-cw to mode locking state. As can be seen here, the pulse energy firstly increased and then experienced an intensity drop before evolving into the stable mode-locking, showing a similar evolution trend to the measured pulse-train in Fig. 3(d). Therefore, the observed phenomena demonstrated that the overall signatures of the soliton booting dynamics in the simulation results are in qualitative agreement with the experiments in both the spectral and time domains.

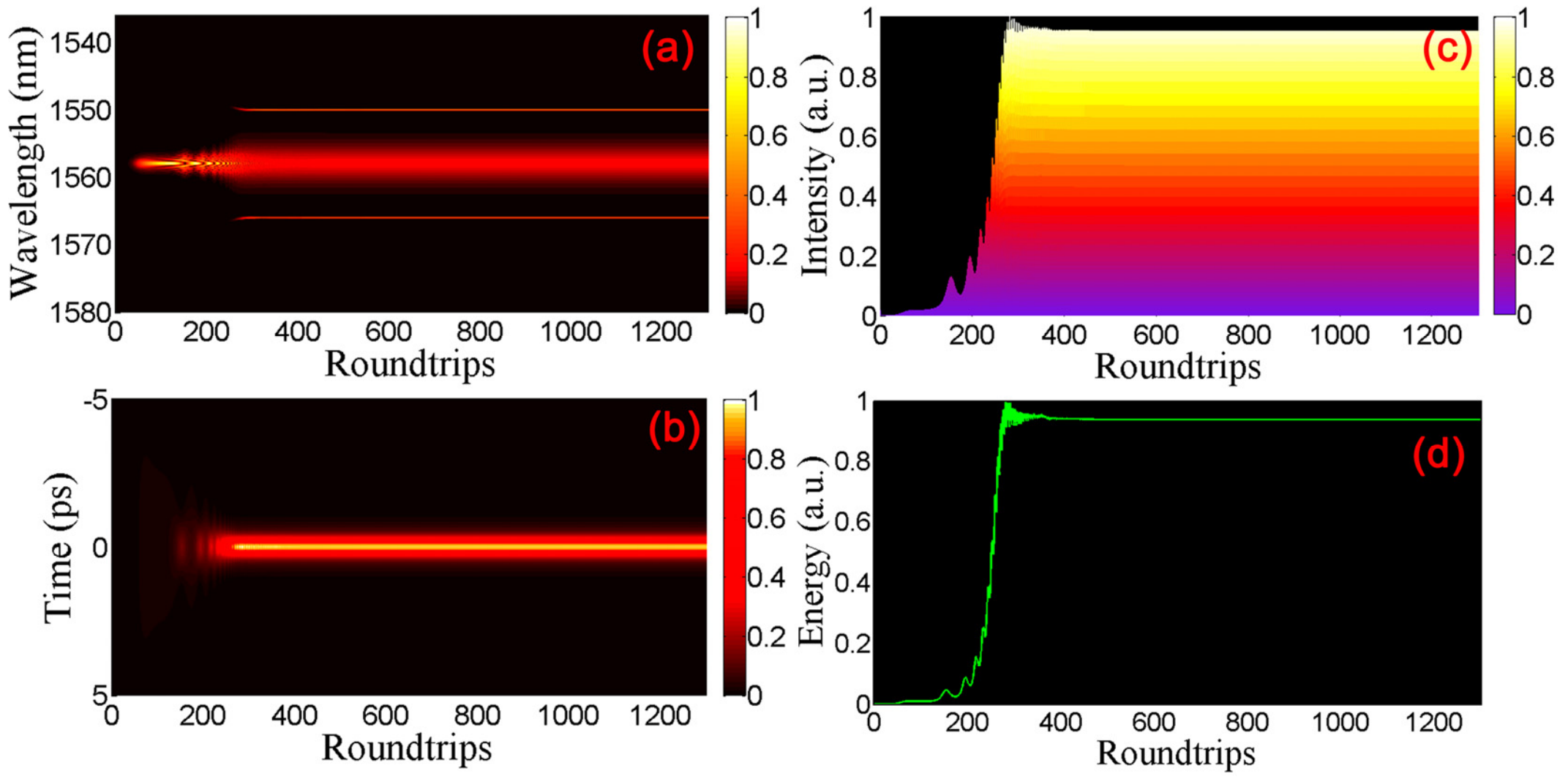

Figure 8. Numerical simulations of conventional soliton booting dynamics in time and spectral domains for 1300 consecutive roundtrips. (a) Spatio-spectral dynamics. (b) Spatio-temporal dynamics. (c) Side view of spatio-temporal dynamics. (d) Pulse energy evolution.

To gain more insight into the simulation results and further verify our experimental results, four transient spectra as well as the corresponding pulse profiles during the soliton booting dynamics at differently representative roundtrips were provided in Fig. 9. It is worth noting that the spectral peaks or dips were alternatively generated in the central part of the pulse spectrum, as shown in the upper row of Fig. 9. In addition, not only these spectral oscillations were obtained, but also the spectral interference patterns were found in simulation. To find the origin of the spectral patterns, we carefully analyzed the pulse profiles corresponding to the spectral states plotted in Fig. 9. As shown in the lower row of Fig. 9, the evolved pulse exhibits evident fine structures at the pedestal of the pulse, and the profile of the pulse pedestal also evolved during the pulse shaping in the cavity, meaning that the transient structural solitons were observed in the simulations[51]. As mentioned above, with the limited bandwidth of the oscilloscope, we cannot resolve the structural profiles of the booting soliton. However, since the spectral dynamics of the soliton booting are well conformed to our experimental results, it was believed that the generated spectral patterns are induced by the transient structured soliton formation during the pulse shaping process. That is, the fine structures of the evolved soliton manifest themselves as the corresponding spectral patterns on the pulse spectrum.

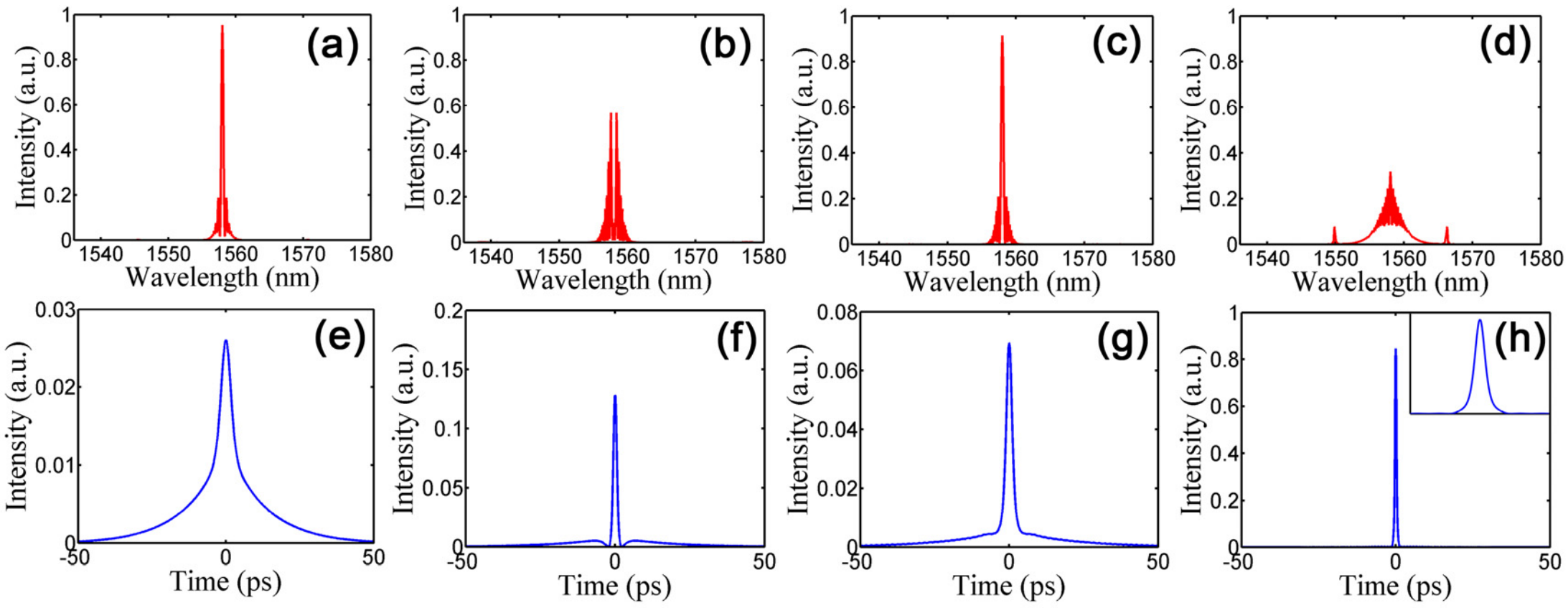

Figure 9. Numerical simulations of four transient spectral and pulse profiles of the booting conventional soliton. (a) and (e) 121 roundtrip. (b) and (f) 154 roundtrip. (c) and (g) 174 roundtrip. (d) and (h) 270 roundtrip.

**Numerical simulations of dissipative soliton booting dynamics in the ultrafast fiber laser**

Then the dissipative soliton booting dynamics in the normal dispersion cavity was simulated by adjusting the fiber dispersion used in the simulation model. Here, the overview of the dissipative soliton booting dynamics was presented firstly by plotting 2700 roundtrips of the soliton buildup evolution both in spectral and temporal domains from the simulation results in Fig. 10. In Fig. 10(a), it can be seen that the laser operated in the cw state firstly, and then the spectral bandwidth broadens as the lightwave propagates in the laser cavity. Note that the sharp-peaked edges were developed during the spectral broadening process, which was in accordance with the experimental results. Figure 10(b) and 10(c) illustrate the pulse evolution over 2700 roundtrips. We can see that the peak intensity of the pulse increased firstly, then decreased sharply until the stable mode locking operation was realized. Then we plotted the pulse energy evolution by integrating the pulse profiles in Fig. 10(d). The pulse energy evolution is still consistent with the experimental results.

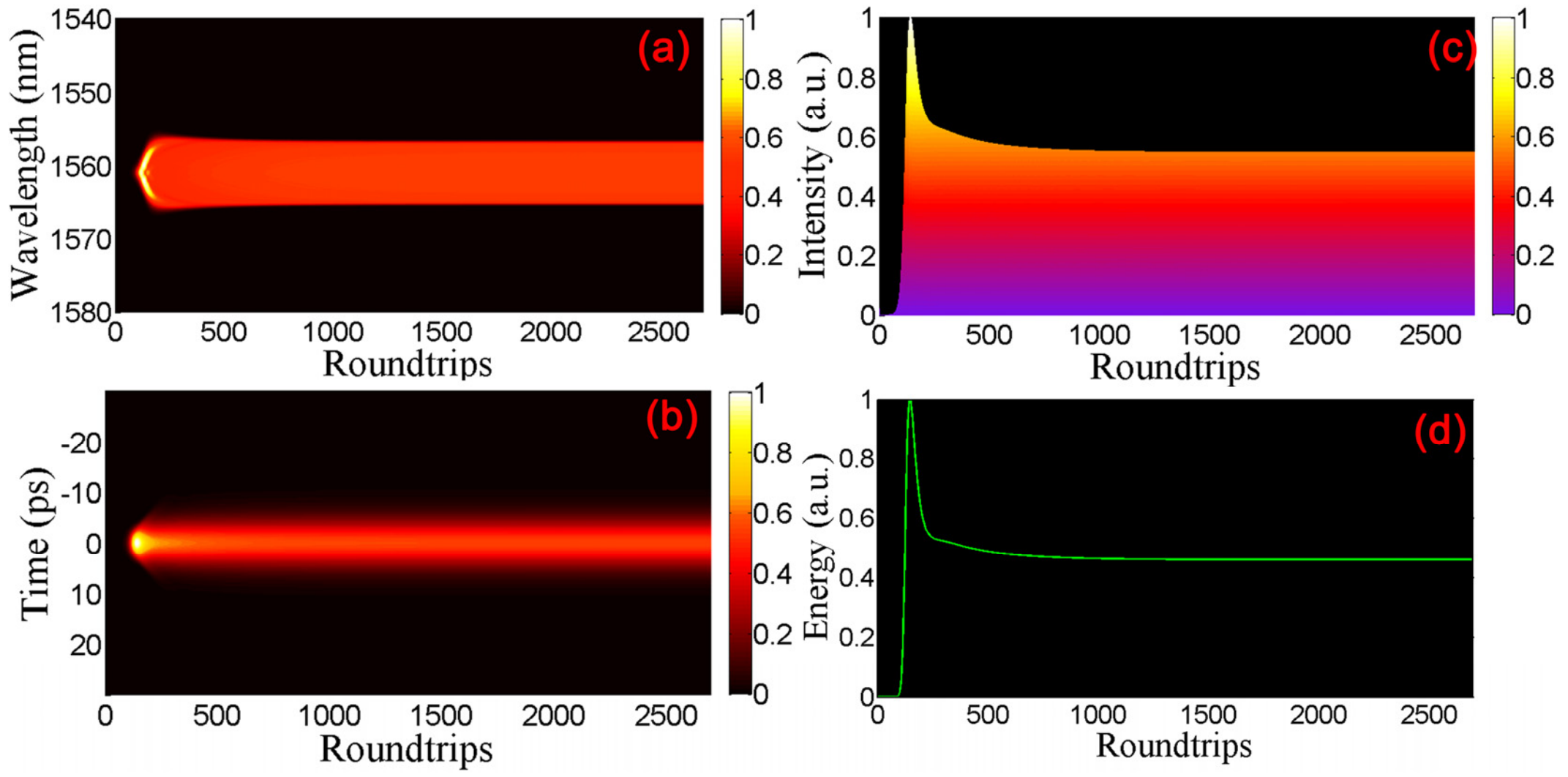

Figure 10. Numerical simulations of dissipative soliton booting dynamics in spectral and time domains for 2700 consecutive roundtrips. (a) Spatio-spectral dynamics. (b) Spatio-temporal dynamics. (c) Side view of spatio-temporal dynamics. (d) Pulse energy evolution.

In the same way, Fig. 11 offered four representative theoretical spectra and the corresponding pulse profiles at roundtrips of 104, 163, 222 and 1700, so as to be in contrast with the experimental results and investigate the details during the soliton booting dynamics. In the frequency domain, from the upper row of Fig. 11 it was found that the spectral patterns appeared on the evolved spectra and the interference depth decreased with the increasing propagation roundtrips, which finally disappeared when realizing the stable mode-locking operation. In addition, the transient shock waves, shown as highest spectral peaks at both sides of spectra with oscillation structures around them, could be observed before the stable mode-locking operation in the simulation results, which is well in agreement with the experimental results. In the time domain, the pulse profiles evolved with cavity roundtrips are shown as the lower row of Fig. 11. Due to the dispersion, the pulse duration broadened with the increasing roundtrips until reaching the limitation of the gain bandwidth and then became stable. Therefore, the saturable absorption effect and the gain bandwidth in the laser cavity would cut off the wings of the pulse and spectrum[11], as illustrated in Fig. 11(f). Moreover, it can be seen that there are also two small humps located at two edges of the evolved soliton in Fig. 11(g). All these features make the dissipative soliton as a structural one in the laser cavity. Note that the fine structures on the dissipative soliton profile in normal dispersion are not as evident as those of conventional soliton booting in anomalous dispersion regime. However, from the simulation results it could be confirmed that the spectral pattern appears on the pulse spectrum when the evolved soliton becomes structural one. Thus, it can be concluded that the evolution characteristics in the spectral domain are dependent on evolution features in the time domain due to the pulse shaping effect. It should be also noted that, due to the different dispersion regime, the two edges of the propagating conventional soliton and dissipative soliton correspond to the central part and both sides of the mode locking spectra[54,55], respectively. We recall that the structures of the solitons during the booting process always appear on both edges of the solitons despite of the soliton types. Thus, the rising fine structures in the soliton profiles would cause the spectral variations (or interferences) most strong in central part for conventional soliton and both edges for dissipative soliton, respectively. This mechanism can be used to explain the difference of spectral evolution in the soliton buildup process between conventional and dissipative solitons in different dispersion regimes.

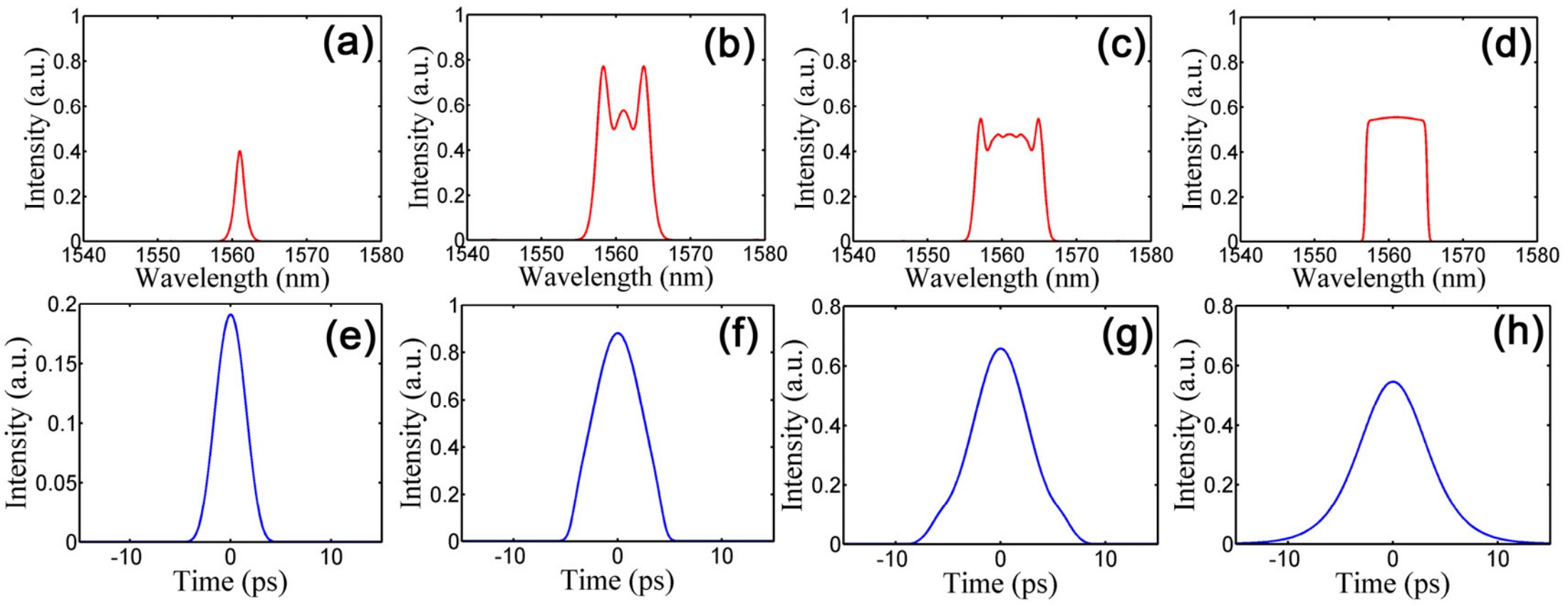

Figure 11. Numerical simulations of four transient spectral and pulse profiles of the booting dissipative soliton. (a) and (e) 104 roundtrip. (b) and (f) 163 roundtrip. (c) and (g) 222 roundtrip. (d) and (h) 1700 roundtrip.

## Discussion

Different from the spectral relaxation oscillation behavior in soliton build up process of the KLM Ti:sapphire lasers recently reported[41], in our experiments the mechanism of spectral relaxation oscillation could be attributed to the transient structural soliton formation during the pulse shaping process. In addition, the spectral evolution characteristics in the anomalous dispersion regime could be concluded as that the highest spectral components with strong relaxation oscillation behavior are always concentrated in the central part of mode-locked soliton during the booting process. While in the normal dispersion regimes, the highest structured spectral spikes spread along both edges for dissipative soliton in normal dispersion regime during the soliton buildup, namely, the transient shock waves could be observed. And no spectral relaxation oscillations were found. The difference of the spectral evolution in the anomalous and normal dispersion regimes could be due to the different pulse shaping mechanisms for both two dispersion regimes. As we know, the front-edge and trailing-edge of the low-chirped conventional solitons correspond to the central frequency of the spectrum, while that of the highly-chirped dissipative solitons correspond to the edge frequencies of the spectrum[54,55]. That might be explained why the strongest spectral patterns aroused by the edge structures of the solitons located at the central part of the mode-locked spectrum for the conventional soliton in the anomalous dispersion regime, while it is at the edge parts for the dissipative soliton in normal dispersion regime, as can be concluded by the reported results above. On the other hand, in this work the soliton booting dynamics in EDF lasers with both the anomalous and net-normal cavity dispersion regime were demonstrated experimentally and theoretically. However, in order to fully investigate the soliton booting dynamics, the fiber lasers at other wavebands or with the all-normal dispersion regime need to be studied in the future to explore whether there exists some other different characteristics in the soliton booting dynamics. Moreover, although the soliton booting dynamics in time domain have not been clearly revealed in our experiment, the problem can be solved by using the time lens measurement[30,56,57] that can be employed to provide a temporal magnification of the evolved pulses. Then the pulse profile in the picosecond or sub-picosecond regime can be measured directly from a high-speed real-time oscilloscope. Nevertheless, since the general spectral evolution tendency of both the experiments and simulations are almost the same, the simulation results of the temporal pulse evolution can still provide a detailed insight into the soliton booting dynamics in the time domain in our work.

In conclusion, we have experimentally and theoretically investigated the soliton booting dynamics in the ultrafast fiber lasers both in the anomalous and net-normal dispersion regimes. Due to the different pulse shaping mechanisms in the anomalous and net-normal dispersion regimes, the soliton booting processes exhibited different characteristics, such as the positions of highest spectral interference pattern and with/without the relaxation oscillation. However, all these spectral evolution dynamics could be explained by the transient structured soliton formation during the pulse shaping from the noise background. The obtained results would pave the way for further investigations of the complex soliton dynamics in nonlinear optical systems, which would be also helpful for the communities dealing with the solitons and fiber lasers.

# Methods

**DFT technique.** When the pulse experiences large group-velocity dispersion (GVD) provided by a dispersive element, spectrum of a pulse is mapped to a temporal waveform whose intensity mimics its spectrum. It means that the spectrum could be captured by a photodetector in a real-time manner. Thus DFT is a powerful method that overcomes the speed limitation of traditional spectrometers and hence enables fast real-time spectroscopic measurements.

**DFT implement.** In our experiment, a ~14 km SMF with a dispersion parameter of 17 ps/km/nm is placed between the laser output and the photodetector, playing the role of dispersive element with large GVD. When the pulse-train comes out from the SMF, the spectrum of each pulse could be observed on the oscilloscope due to the DFT technique. Therefore the spectral evolution in the soliton booting process could be measured in a real-time diagnostic method.

## Acknowledgments

We acknowledge financial support from the National Natural Science Foundation of China (NSFC) (11304101, 11474108, 61307058, 61378036); Guangdong Natural Science Funds for Distinguished Young Scholar (2014A030306019); Program for the Outstanding Innovative Young Talents of Guangdong Province (2014TQ01X220); Program for Outstanding Young Teachers in Guangdong Higher Education Institutes (YQ2015051); Science and Technology Project of Guangdong (2016B090925004).

## Additional information

**Competing financial interests:** The authors declare no competing financial interests.

# Unveiling soliton booting dynamics in ultrafast fiber lasers: supplementary material

**This document provides supplementary information to “Unveiling soliton booting dynamics in ultrafast fiber lasers”. The supplementary material contains three sections, as shown in the following: in the first section, the experimental phenomena of the relaxation oscillations for the temporal pulse occurring during soliton booting process were provided in a large time range both in the anomalous and net-normal dispersion regime. In the following, we provided six supplementary movies to show the spectral and temporal dynamics of the mode-locking transition experimentally and theoretically for both the anomalous and net-normal dispersion regimes. In the last section, we provide the simulation laser configuration, mathematical model, and the parameters used for simulation.**

## 1. Relaxation oscillation behavior of the soliton booting process in the time domain.

When the mode locking operation of the ultrafast fiber laser was periodically initiated (or stopped) by the mechanical chopper, we respectively measured the pulse-trains in a large time range with a few microseconds in the anomalous and net-normal dispersion regimes of our fiber lasers by the oscilloscope. In this case, both the pulse-trains operating in anomalous and normal dispersion regimes show the strong relaxation oscillation before achieving the stable mode-locking state. The oscillation behavior is very similar to the Q-switched mode locking operation in ultrafast fiber lasers. The results are plotted in Fig. S1. As can be seen from Fig. S1, despite of the dispersion regime, the pulse relaxation oscillation could be always observed in the ultrafast fiber lasers during the soliton booting process. However, the soliton booting time in normal dispersion regime is shorter than that in the anomalous dispersion regime. The booting time of the mode-locked soliton has been demonstrated to be related to the intracavity pulse energy[1]. Note that the pulse energy of dissipative soliton is higher than that of conventional soliton in anomalous dispersion regime in our fiber lasers because of different pump power level set in the experiments. Therefore, it could be expected that the soliton buildup time of conventional soliton would be longer than the dissipative soliton, as depicted in Fig. S1.

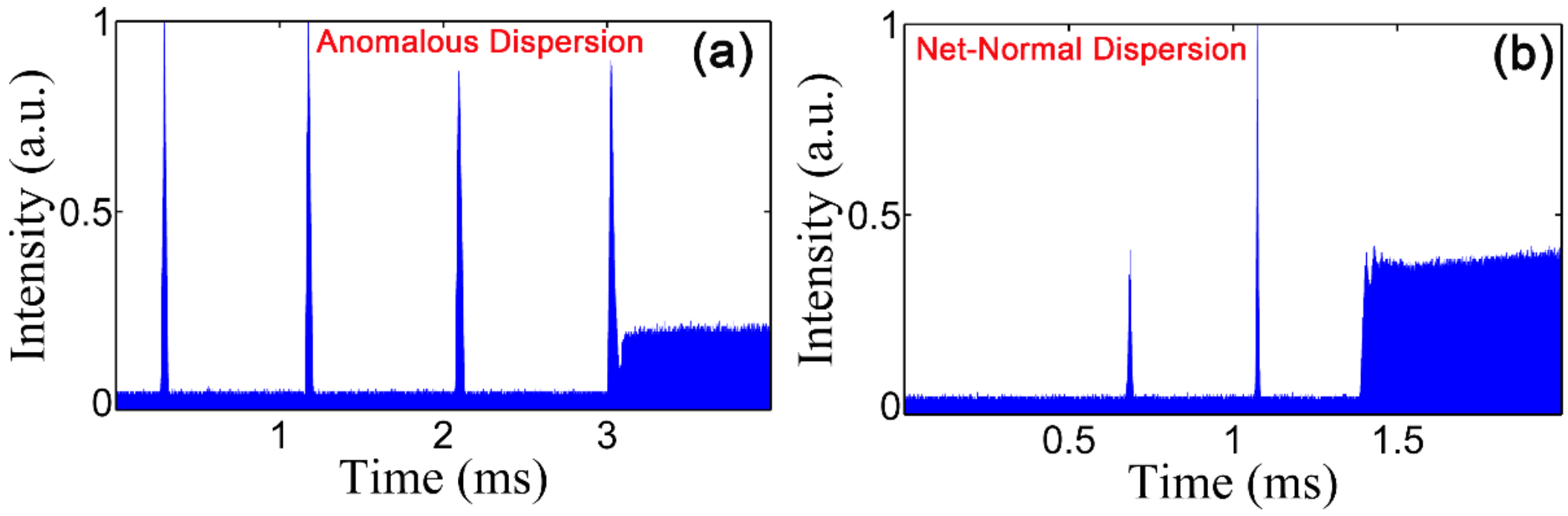

Figure S1. Measured pulse-trains of soliton booting process: (a) in the anomalous dispersion regime; (b) in the net-normal dispersion regime.

**2. Supplementary movies.**

**2.1. Experimental observation of the spectral dynamics in ultrafast fiber laser**

Two videos showing the spectral evolution for both the conventional and dissipative solitons were presented here. Two frames of the spectral evolution extracted from the movies are shown in Fig. S2(a) and S2(b), which represent the anomalous and the normal dispersion regimes, respectively. The Movie 1 illustrates the real-time spectral evolution of the mode-locking transition in the anomalous dispersion regime, from which can be seen clearly the oscillation structure of the evolved spectrum before the onset of mode locking. The Movie 2 exhibits the spectral dynamics of the dissipative soliton evolution in ultrafast fiber laser operating in the net-normal dispersion regime, showing that the spectral peaks could be observed on both edges of the spectrum in the net-normal dispersion regime. Here, it should be noted that the movies are all extracted from the experimental data.

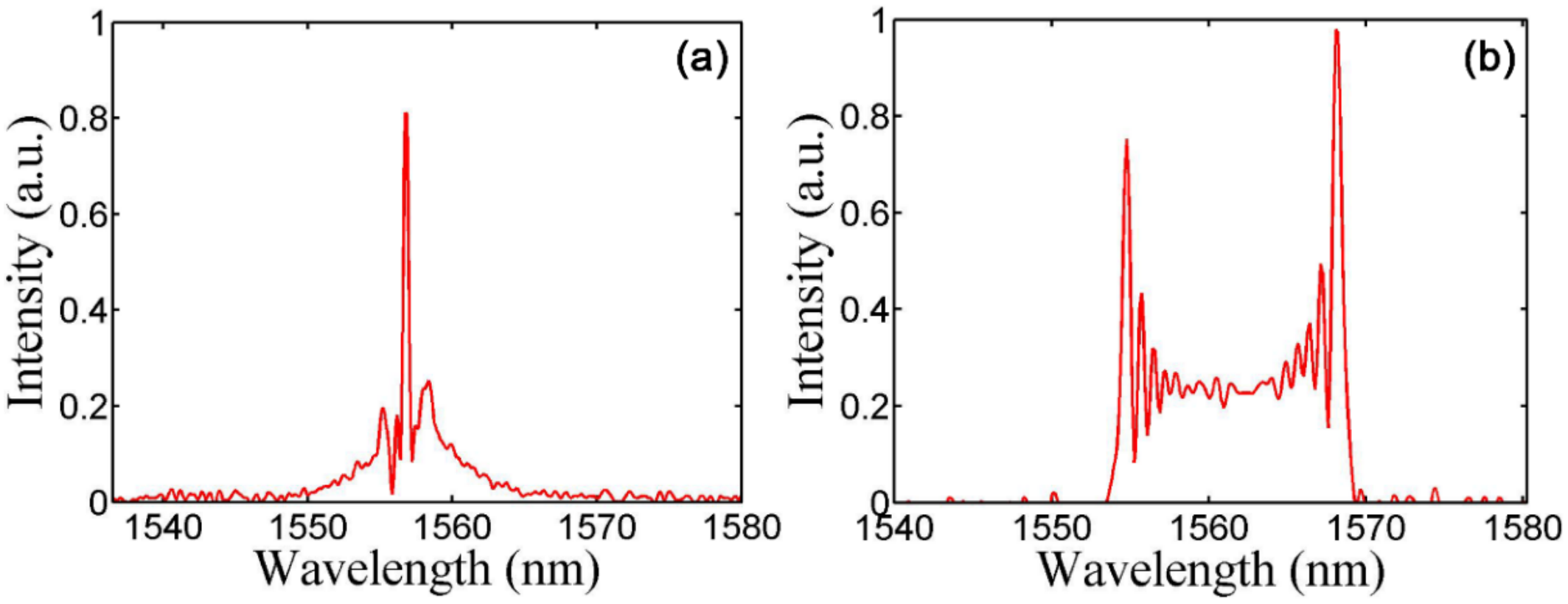

Figure S2. Two frames of the spectral evolutions extracted from the Movies 1 and 2. (a) A spectrum in the spectral evolution of conventional soliton; (b) A spectrum in the spectral evolution of dissipative soliton.

### 2.2. Simulation results of the mode-locking dynamics in ultrafast fiber lasers

Because the pulse evolution during the soliton booting process in ultrafast fiber lasers could not be resolved experimentally, we would like to show the reconstructed simulation data of the temporal dynamics for the mode locking as a movie here. For better clarity, Fig. S3(a) and S3(b) display two frames of supplementary movies of the pulse evolution corresponding to Movies 3 and 4, showing the pulse propagation in the laser cavity in the anomalous and normal dispersion regimes, respectively. As can be seen clearly from Movies 3 and 4, the structural solitons were formed during the pulse shaping process from noise background to stable mode-locked soliton for both dispersion regimes. In addition, we also show the reconstructed simulation data of the spectral dynamics for laser mode locking in the Movies 5 and 6, which correspond to the anomalous and normal dispersion regimes, respectively. Meanwhile, two frames of supplementary movies of the spectral evolution corresponding to Movies 5 and 6 are depicted in the Fig. S4.

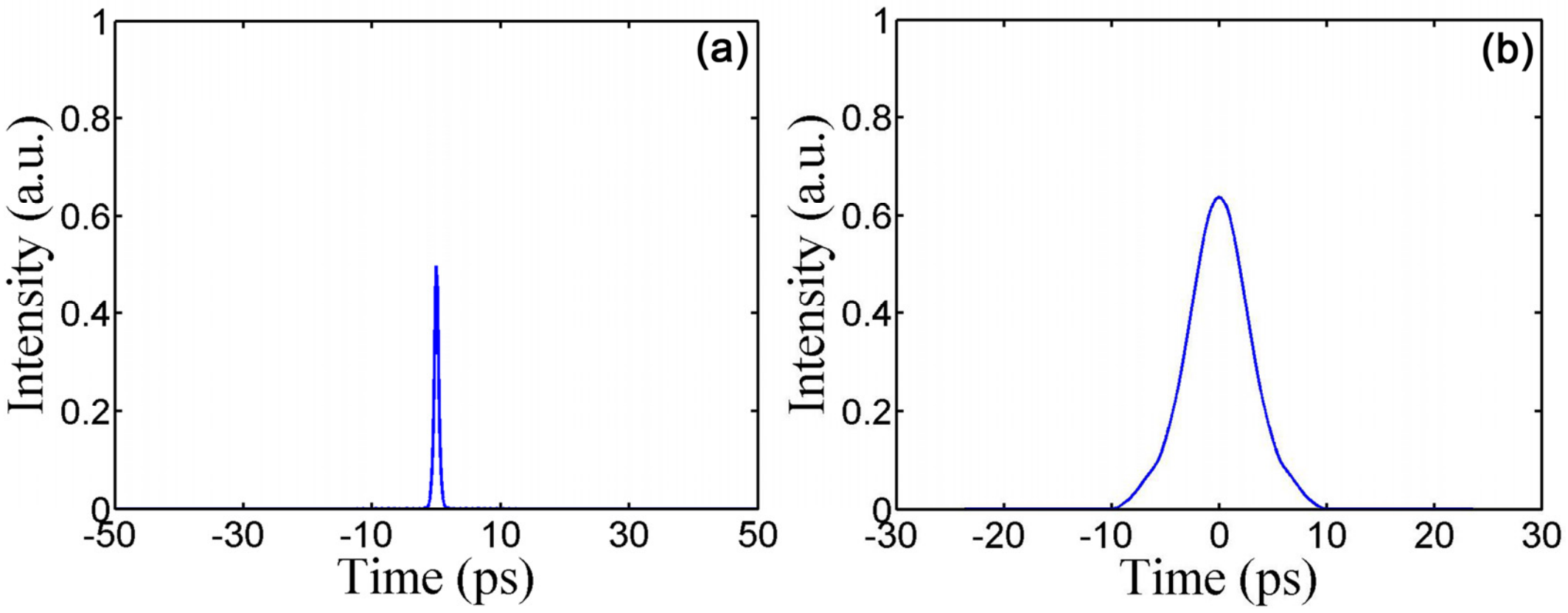

Figure S3. Two frames of the pulse evolutions extracted from the Movies 3 and 4. (a) An evolving pulse profile of the conventional soliton; (b) An evolving pulse profile of the dissipative soliton.

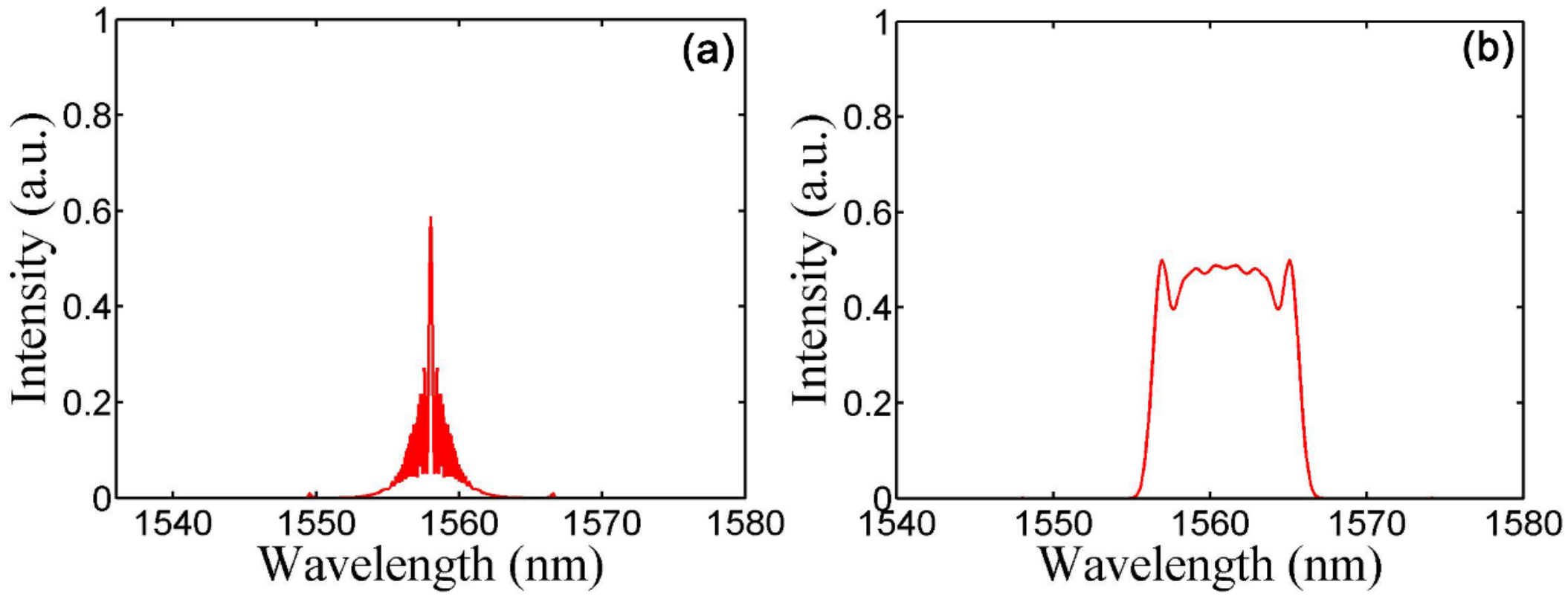

Figure S4. Two frames of the spectral evolutions extracted from the Movies 5 and 6. (a) Conventional soliton; (b) Dissipative soliton.

### 3. **Laser setup and numerical model**

The fiber ring laser cavity that we study is depicted in Fig. S5, which is a conceptual one but generally enough for us to investigate the soliton booting dynamics of passively mode-locked lasers. The ring cavity consists of the active Er-doped fiber (EDF) with variable length, single mode fiber (SMF) as the passive fiber, a saturable absorber (SA) and an output coupler (OC).

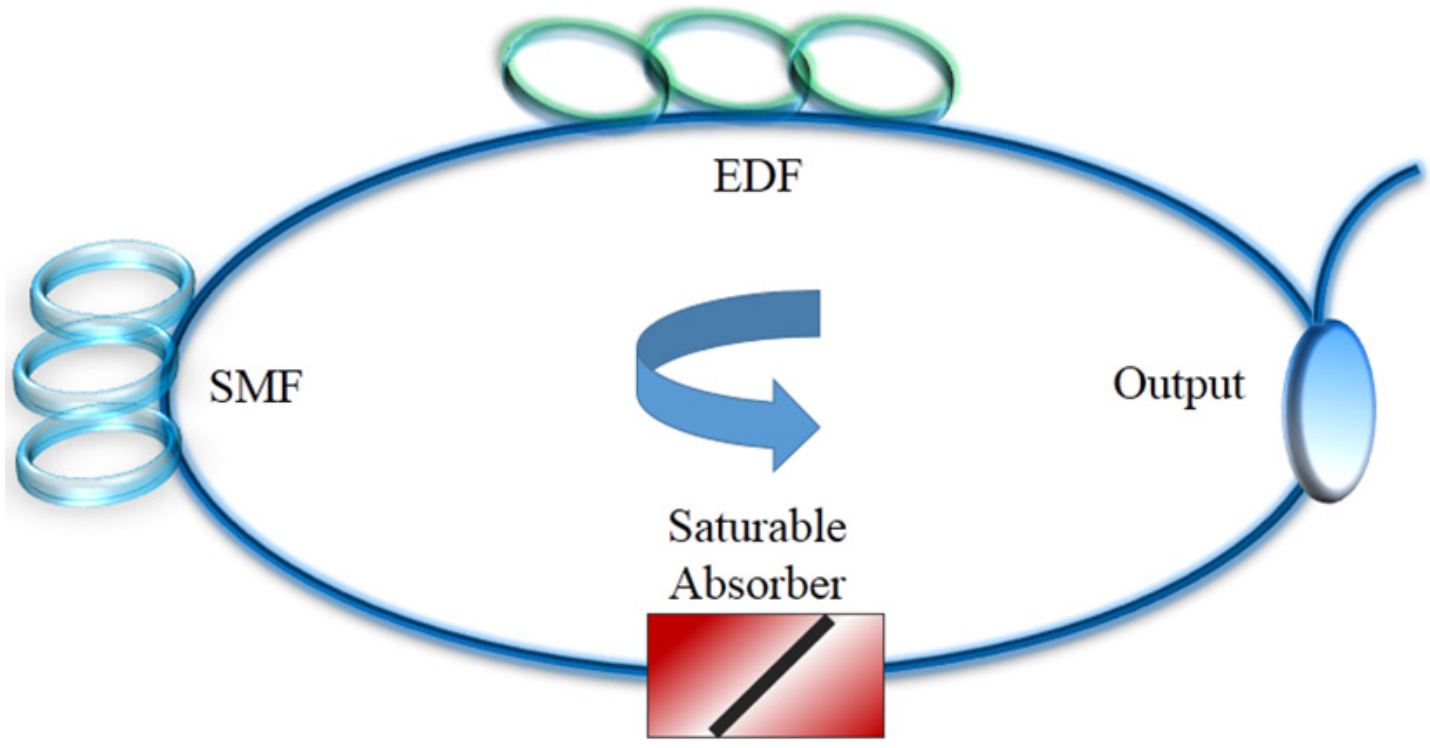

Figure S5. Schematic of the fiber laser cavity for the simulations.

We numerically study the starting dynamics of a soliton in the fiber laser cavity both in the anomalous and normal dispersion regimes. Numerical simulations are based on an extended nonlinear Schrodinger equation, which is solved with a standard symmetric split-step algorithm for both segments of the active and passive fibers. The modeling includes the physical terms such as the self-phase modulation, the group velocity dispersion of fiber, and the saturated gain with a finite bandwidth:

$$\frac{\partial A}{\partial z}+i\frac{\beta_2}{2}\frac{\partial^2 A}{\partial t^2}=\frac{g}{2A}+i\gamma|A|^2 A+\frac{g}{2\Omega_g^2}\frac{\partial^2 A}{\partial t^2} \tag{1}$$

Here, $A$ is the slowly varying amplitude of the pulse envelope, the variable $z$ and $t$ represent the propagation distance and the time, respectively. $\beta_2$ and $\gamma$ are denoted as the fiber dispersion and the cubic refractive nonlinearity of the fiber, respectively. The bandwidth of the gain spectrum is $\Omega_g$, $g$ describes the gain function of the EDF which could be given by:

$$g=g_0*\exp(-E_P/E_{sat})$$

Where $g_0$ is the small-signal gain, and $E_P=\int|A|^2 dt$ is the pulse energy, $E_{sat}$ is the gain saturation energy which relies on pump power. The SA is modelled by a simplified transfer function[2]:

$$T(t)=1-\frac{q_0}{1+|A(t)|^2/P_{sat}}$$

Here, $q_0$ is the modulation depth, $|A|^2$ the instaneous pulse power and $P_{sat}$ the saturation power for SA.

The initial conditions used in our numerical simulation are a 1 ps time-bandwidth limited sech²-shaped pulse with 0.01 W of peak power in the anomalous dispersion regime and a 1 ps time-bandwidth limited sech²-shaped pulse with 0.00001 W of peak power in the normal dispersion regime, respectively. Both the phases of the initial pulse used in the anomalous and normal dispersion regimes are zero. This method can generally accelerate the convergence of the simulation. When the pulse evolves in the cavity, after a cavity roundtrip, the result of calculation of pulse is then used as a new initial signal in the next roundtrip calculation.

**Table S1. Simulation parameters of the fiber ring laser in the anomalous and normal dispersion regimes.**

| Dispersion regime | | Anomalous dispersion | Normal dispersion |
|---|---|---|---|
| Element | Parameter | Value | |
| EDF | Length $L_{EDF}$<br>2nd order dispersion $\beta_2$<br>Nonlinear parameter $\gamma$<br>Gain bandwidth $\Omega_g$<br>Small signal gain $g_0$<br>Saturation energy $E_{sat}$ | 4. 1 m<br>-0.009 $\mathrm{ps}^2/\mathrm{m}$<br>0.003 1/W/m<br>50 nm<br>0.205 dB/m<br>0.15 J | 10.14 m<br>0.0553 $\mathrm{ps}^2/\mathrm{m}$<br>0.003 1/W/m<br>30 nm<br>0.1395 dB/m<br>0.15 J |
| SMF | Length $L_{SMF}$<br>2nd order dispersion $\beta_2$<br>Nonlinear parameter $\gamma$ | 11.57 m<br>-0.0209 $\mathrm{ps}^2/\mathrm{m}$<br>0.003/W/m | 12.027 m<br>-0.022 $\mathrm{ps}^2/\mathrm{m}$<br>0.003/W/m |
| SA | Modulation depth $q_0$<br>Saturation power $P_{sat}$ | 0.451<br>450 W | 0.7<br>470 W |
| OC | Out-coupling parameter $R_{out}$ | 10% | 10% |

We numerically simulated the passive mode locking operation both in the anomalous and normal dispersion regimes and the values of initial pulses used are described above. Here, it is worth noting that the small-signal gain $g$ represents the pump power, and the small-signal gain $g_0$ only needs to be considered in the gain medium of EDF. Therefore, when the evolved pulse passes through the EDF, the pulse energy will increase. Note that one cavity roundtrip includes the action of the active fiber, passive fiber, output coupler and SA. For better clarity, all the simulation parameters are shown in Table S1 above.